\documentclass{article}
\usepackage{spconf,amsmath,graphicx,hyperref}
\usepackage{graphics}
\usepackage{subfig}
\usepackage{caption}
\usepackage{changepage}
\usepackage{booktabs}
\usepackage{mathtools}
\usepackage{threeparttable}
\usepackage{multirow}

\makeatletter
\def\UrlAlphabet{
    \do\a\do\b\do\c\do\d\do\e\do\f\do\h\do\i\do\j%
    \do\k\do\l\do\m\do\q\do\r\do\s\do\t%
    \do\u\do\v\do\w\do\x\do\y\do\z}
\def\UrlDigits{\do\1\do\2\do\3\do\4\do\5\do\6\do\7\do\8\do\9\do\0}
\g@addto@macro{\UrlBreaks}{\UrlOrds}
\g@addto@macro{\UrlBreaks}{\UrlAlphabet}
\g@addto@macro{\UrlBreaks}{\UrlDigits}
\makeatother

\def\x{{\mathbf x}}

\newcommand{\eg}{{\emph{e.g.}}}

\title{Subjective Evaluation of Frame Rate in Bitrate-Constrained \\Live Streaming}

\name{Jiaqi He\textsuperscript{1}, 
Zhengfang Duanmu\textsuperscript{2}\sthanks{While Zhengfang is currently employed by Netflix, this research was completed as his individual work. Netflix had no ideating, funding, or supporting this research.}, and
Kede Ma\textsuperscript{1}}
\address{\textsuperscript{1}Department of Computer Science, City University of Hong Kong\\
\textsuperscript{2}Netflix Inc.} 
\begin{document}

\maketitle

\begin{abstract}
Bandwidth constraints in live streaming require video codecs to balance compression strength and frame rate, yet the perceptual consequences of this trade-off remain underexplored.
We present the high frame rate live streaming (HFR-LS) dataset, comprising $384$ subject-rated 1080p videos encoded at multiple target bitrates by systematically varying compression strength and frame rate.
A single-stimulus, hidden-reference subjective study shows that frame rate has a noticeable effect on perceived quality, and interacts with both bitrate and source content. The HFR-LS dataset is available at \url{https://github.com/real-hjq/HFR-LS} to facilitate research on bitrate-constrained live streaming.
\end{abstract}


\section{Introduction}
\label{sec:intro}

Interactive video applications such as cloud gaming~\cite{huang2013gaminganywhere}, virtual reality~\cite{guan2019pano}, and high-resolution video conferencing~\cite{holub2012ultragrid} are accelerating the demand for live streaming that delivers both low latency and perceptually high quality. Meeting this demand is challenging under bandwidth constraints. Encoders must continually balance spatial fidelity and temporal smoothness by jointly adjusting compression strength, spatial resolution, and frame rate~\cite{ISO2012Dash}. Increasing frame rate generally improves motion representations, but also spreads the bitrate budget over more frames, intensifying pre-frame compression and potentially degrading spatial detail.

Prior high frame rate (HFR) datasets (\eg, Waterloo-IVC HFR~\cite{nasiri2015hfr}, BVI-HFR~\cite{mackin2015study}, UVG~\cite{mercat2020uvg}, LIVE-YT-HFR~\cite{pavan2021HFR}, and ETRI-LIVE STSVQ~\cite{lee2022etri})  vary quantization parameter (QP)/constant rate factor (CRF), spatial resolution, and frame rate \textit{independently}. Such designs are misaligned with bitrate-constrained live streaming, where the key choice is allocating a \textit{fixed} target bitrate between compression strength, spatial resolution, and frame rate. Consequently, the perceptual trade-offs between these factors in realistic live-streaming settings remain insufficiently understood.

We address this gap with the high frame rate live streaming (HFR-LS) dataset and a controlled subjective study designed around bitrate-constrained streaming. HFR-LS comprises $32$ high-quality source sequences at $120$ fps resized to 1080p, each encoded into $12$ representations spanned by four target bitrates and three frame rates, yielding $384$ processed videos. A single-stimulus, hidden-reference protocol~\cite{bt500subjective} under standardized viewing conditions is adopted to collect difference mean opinion scores (DMOSs).

Analysis of HFR-LS reveals frame rate as a key contributor to perceived quality, interacting with both bitrate and source content. To further situate these insights, we benchmark a range of full-reference and no-reference video quality assessment (VQA) models on HFR-LS, underscoring the need for improved quality predictors that explicitly account for bitrate-aware frame-rate adaptation.

\begin{figure*}[t]
    \captionsetup{font=small}
    \centering
    \includegraphics[width=\textwidth]{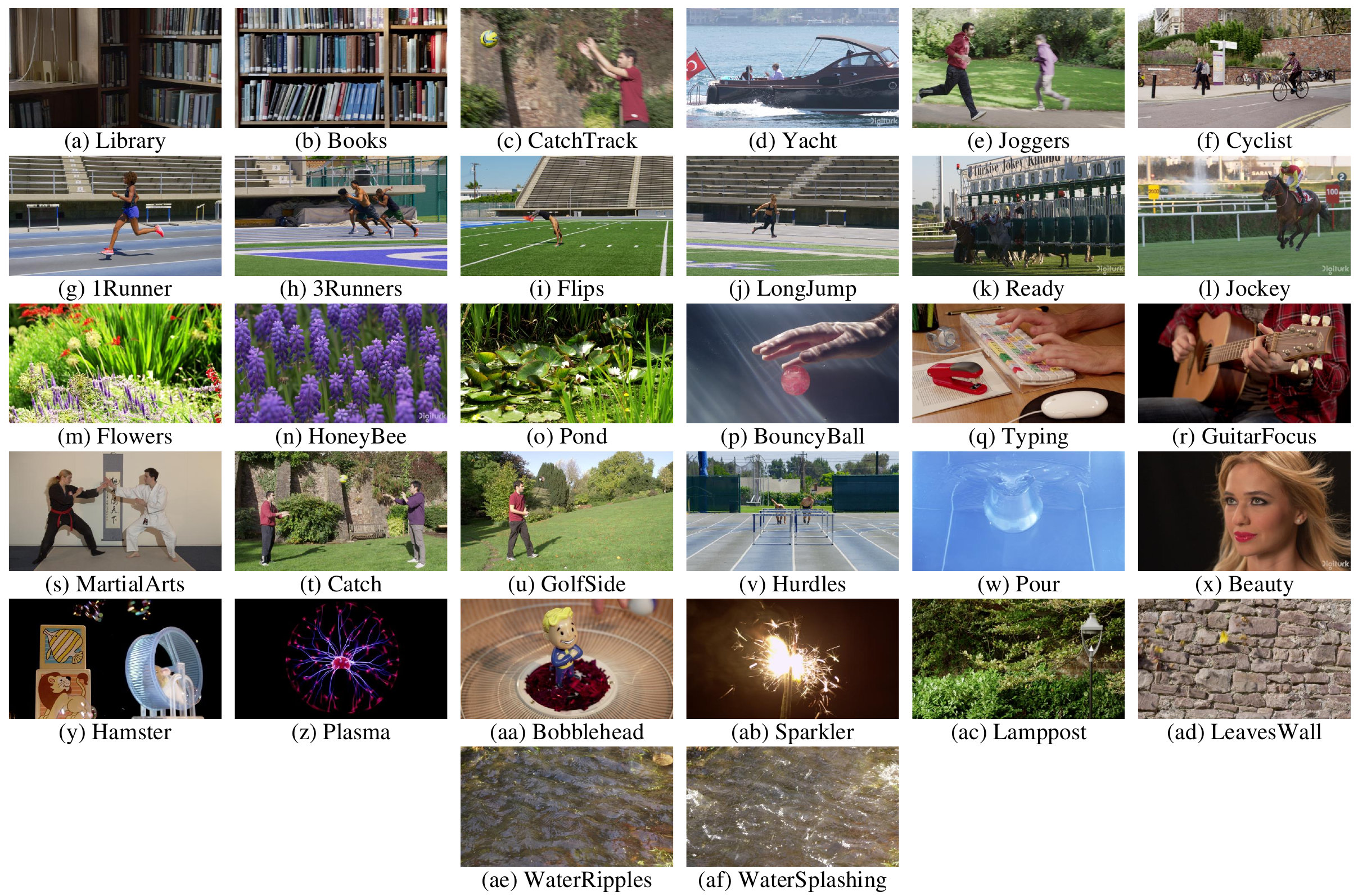}
    \caption{Sample frames from the proposed HFR-LS dataset, among which (a)-(l) were captured with camera motion.}\label{fig:sample_frames}
\end{figure*}

\section{Subjective User Study}
In this section, we design a subjective user study to quantify the perceptual trade-offs between compression strength and frame rate under typical bitrate constraints encountered in live streaming scenarios.

\subsection{Dataset Construction}

\begin{figure}
    \captionsetup{font=small}
    \centering
    \includegraphics[width=0.42\textwidth]{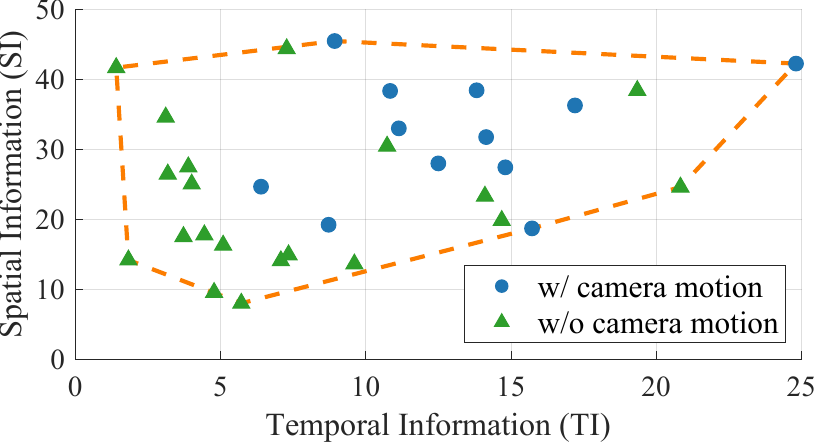}
    \caption{Scatter plot of SI vs. TI for HFR-LS.}\label{fig:scatter_siti}
\end{figure}

\noindent\textbf{Source Content.} We curated a set of $32$ high-quality videos, each originally captured at $120$ fps in accordance with ITU-R BT. 500 guidelines~\cite{bt500subjective}: $22$ from BVI-HFR~\cite{mackin2015study}, $5$ from  UVG~\cite{mercat2020uvg}, and  $5$ from LIVE-YT-HFR~\cite{pavan2021HFR}.
All sources were converted to $1,920 \times 1,080$, YUV 4:2:0, 8-bit format, and trimmed to $5$ seconds. $12$ videos contain camera motion; the remaining contain static-camera content, spanning a broad range of dynamics (see Fig.~\ref{fig:sample_frames}). To characterize content diversity, we computed spatial information (SI) and temporal information (TI) for each source (per ITU-T P.910), confirming coverage from low to high motion/detail (see Fig.~\ref{fig:scatter_siti}).

\noindent\textbf{Encoded Representations.} Each source was encoded into $12$ representations using H.264 at a fixed resolution (1080p), crossing $4$ target bitrates ($5$, $7$, $10$, and $15$ Mbps) with $3$ frame rates ($30$, $60$, and $120$ fps). These operating points reflect common live streaming ladders recommended by major platforms (see Table~\ref{tab:encoding_ladder}), prioritizing bitrate budgeting rather than QP/CRF control. We used FFmpeg~\cite{ffmpeg} (preset: \textit{fast}, tune: \textit{zerolatency}) to encode these videos, and applied frame dropping for frame rate down-conversion\footnote{The complete encoding command is \texttt{ffmpeg -f rawvideo -s:v 1920x1080 -r 120 -pix\text{\textunderscore}fmt yuv420p -i input.yuv -r fps -c:v libx264 -preset fast -tune zerolatency -b:v bitrate -an output.mp4}.}. The resulting HFR-LS dataset contains a total of  $384$ processed video clips.

\begin{table}[t]
\captionsetup{font=small}
\small
\centering
\caption{Video encoding ladder.}
\begin{tabular}{ c | c c | c | c c }
\toprule
     \# & Bitrate & Frame rate & \# & Bitrate & Frame rate \\ [0.5ex] 
\hline
    1 & 5.0 Mbps & 30 fps & 7 & 10.0 Mbps & 30 fps\\ 
    2 & 5.0 Mbps & 60 fps & 8 & 10.0 Mbps & 60 fps\\
    3 & 5.0 Mbps & 120 fps & 9 & 10.0 Mbps & 120 fps\\
    4 & 7.0 Mbps & 30 fps & 10 & 15.0 Mbps & 30 fps\\
    5 & 7.0 Mbps & 60 fps & 11 & 15.0 Mbps & 60 fps\\
    6 & 7.0 Mbps & 120 fps & 12 & 15.0 Mbps & 120 fps\\
\bottomrule
\end{tabular}
\label{tab:encoding_ladder}
\end{table}

\subsection{Subjective Study}
\noindent\textbf{Protocol and Environment.} We employed a single stimulus, hidden-reference paradigm~\cite{bt500subjective}, where each video was rated independently on a $0$-$100$ quality scale. Source reference videos were included to compute DMOSs, helping mitigate potential content preference bias. All videos were displayed at their native resolution on a Truecolor ($32$-bit), $120$ Hz monitor, calibrated to ITU-R BT.500 standards~\cite{bt500subjective}. A graphical user interface was used to display the videos in random temporal order, and participants recorded subjective ratings during this process.

The testing was conducted in an indoor office setting under controlled lighting conditions to minimize external visual disturbances. Participants were seated approximately $1.5$ times the screen height from the monitor, ensuring a standard viewing distance. $30$ na\"{i}ve subjects ($12$ females and $18$ males, aged $20$-$30$) participated in the study. Prior to testing, each participant’s visual acuity and color vision were confirmed. A training session was conducted to familiarize participants with the testing procedure. The test material totaled $2,080$ seconds in total ($32$ sources $\times$ $13$ representations $\times$ $5$ seconds), which was administered in two sessions separated by a $5$-minute break to reduce fatigue. Participants took approximately one hour, on average, to complete the test.

\begin{figure}[t]
    \captionsetup{font=small}
    \centering
    \subfloat[]{\includegraphics[width=0.49\linewidth]{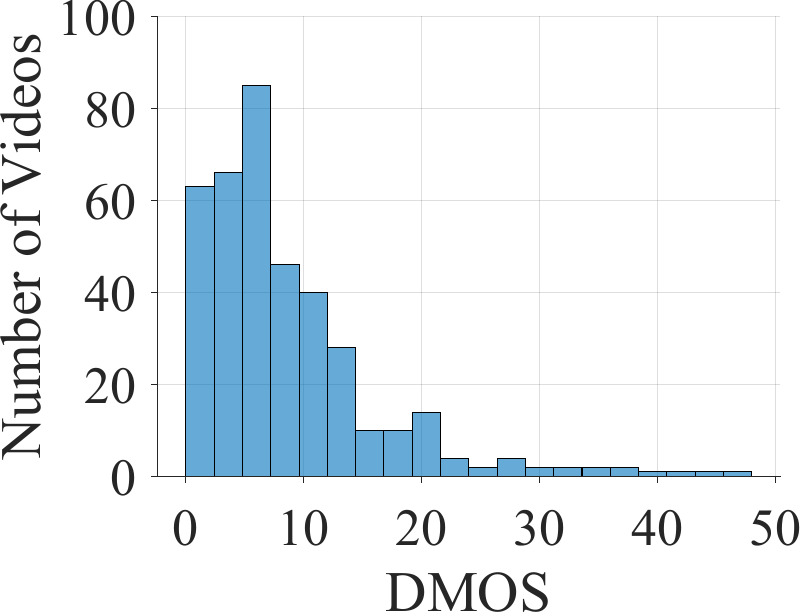}\label{fig:histogram}}\hspace{0.5mm}
    \subfloat[]
    {\includegraphics[width=0.49\linewidth]{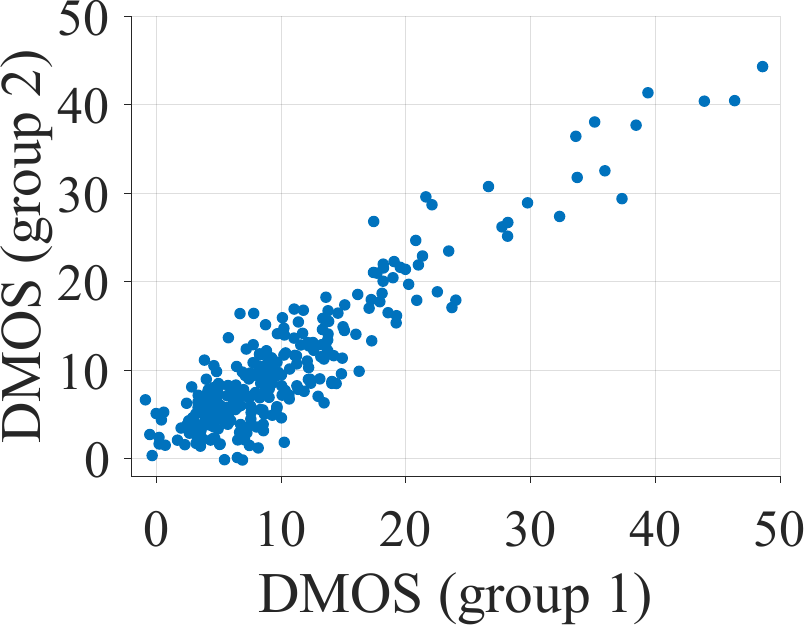}\label{fig:scatter_dmos}}
    \caption{(a) Histogram of DMOSs in $20$ equally spaced bins. (b) Scatter plot of DMOSs between two equal, disjoint groups of subjects.}
\end{figure}

\noindent\textbf{Data Processing.} The raw subjective ratings were normalized by computing per-subject Z-scores, resulting in a matrix $\mathbf{Z}$, where each entry $z_{ij}$ represents the Z-score assigned by subject $i$ to video $j$. We then applied the outlier removal method in~\cite{bt500subjective}, and excluded four invalid participants. Finally, the retained Z-scores were linearly mapped to the $[0,100]$ range using the following transformation:
\begin{equation}
    z^{\prime}_{ij} = \frac{100(z_{ij}+6)}{9}.
\end{equation}
The MOS for each processed video was computed by averaging the rescaled Z-scores across all valid subjects.
The DMOS was then derived by subtracting the MOS of each processed video from that of its corresponding reference video.

Fig.~\ref{fig:histogram} shows the histogram of the resulting DMOSs, ranging from $0$ to $47$ with a mean of $8.57$ and a standard deviation of $7.64$. To assess inter-subject consistency, we randomly divided the subjects into two equal, disjoint groups, and computed the Pearson linear correlation coefficient (PLCC) between the corresponding DMOSs. Across $100$ random splits, the median PLCC is $0.91$, demonstrating high inter-subject consistency (see also Fig.~\ref{fig:scatter_dmos}).

\section{Subjective Data Analysis}
This section addresses two primary questions: (1) Is there a perceptual trade-off between compression strength and frame rate in live streaming? (2) If such a trade-off exists, in which specific scenarios does it manifest?

\begin{figure*}[t]
    \captionsetup{font=small}
    \centering
    \subfloat[w/ camera motion \& $\text{TI} > 6$]{\includegraphics[width=0.32\linewidth]{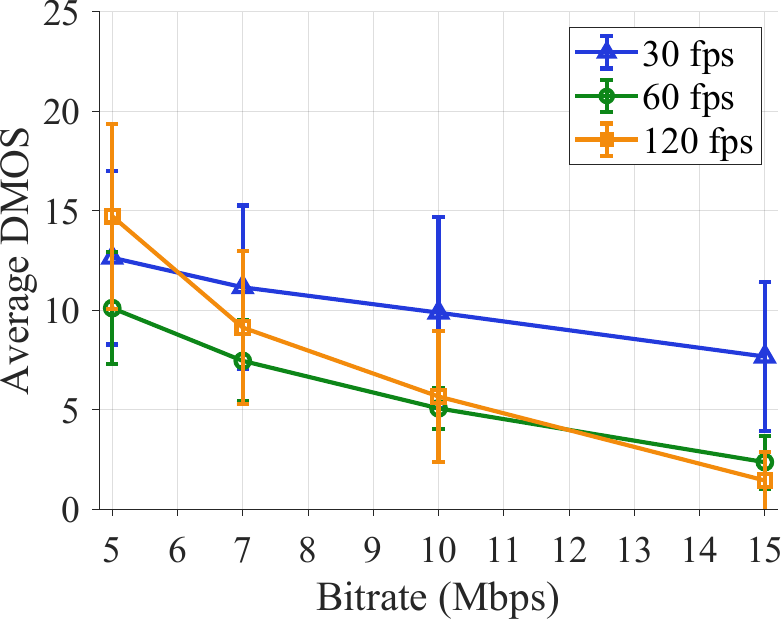}\label{fig:dmos_bitrate1}}\hspace{1mm}
    \subfloat[w/o camera motion \& $\text{TI} > 6$]
    {\includegraphics[width=0.32\linewidth]{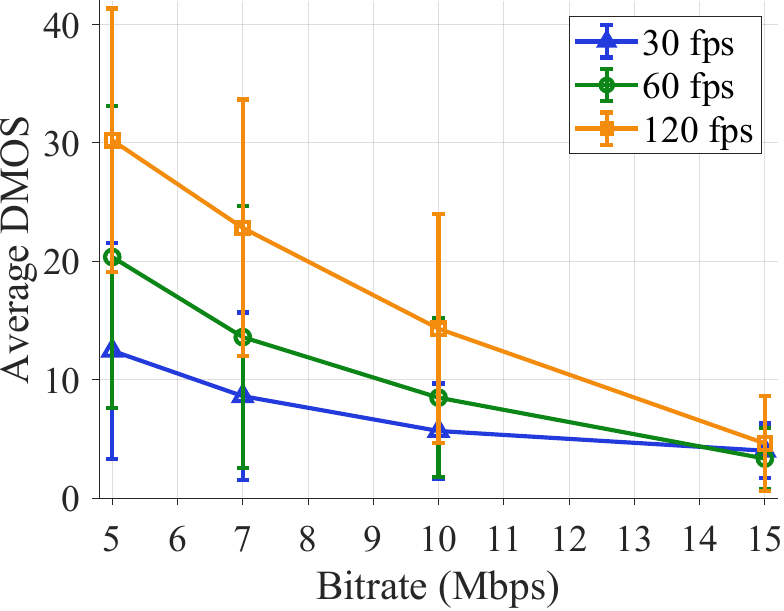}\label{fig:dmos_bitrate3}}\hspace{1mm}
    \subfloat[w/o camera motion \& $\text{TI} < 6$]
    {\includegraphics[width=0.32\linewidth]{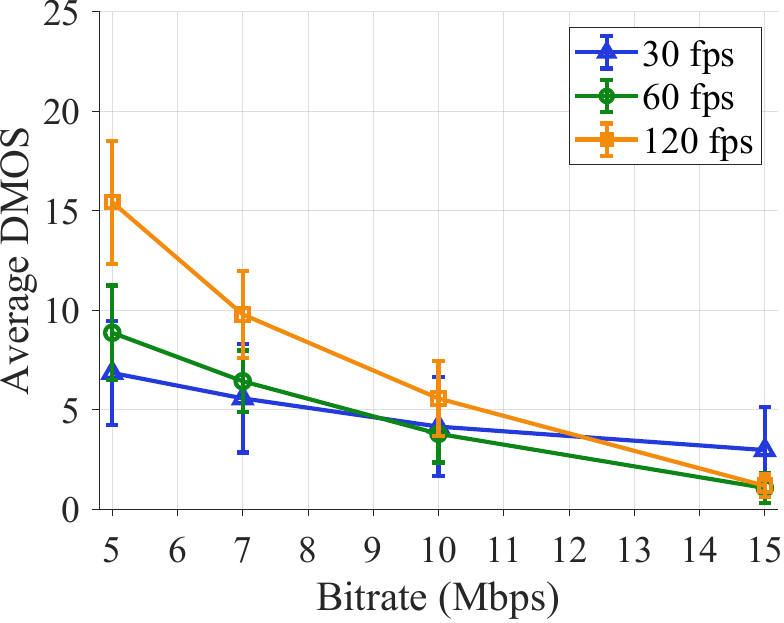}\label{fig:dmos_bitrate2}}
    \caption{Average DMOS as a function of bitrate for different frame rates and source content: (a) $12$ source videos with camera motion ($\text{TI} > 6$), (b) $9$ source videos with significant temporal variations ($\text{TI} > 6$) and no camera motion, and (c) $11$ source videos with slight temporal variations ($\text{TI} < 6$) and no camera motion.  Error bars indicate the standard errors.}
    \label{fig:dmos_bitrate}
\end{figure*}

\subsection{Qualitative Analysis}
To illustrate the trade-off between compression strength and frame rate in live streaming, we present line plots of the average DMOS against bitrate for different frame rates and source content in Fig.~\ref{fig:dmos_bitrate}. 

Our results indicate that perceived video quality generally improves with increasing bitrates. At lower bitrates (\eg, $5$ Mbps), videos encoded at higher frame rates (\eg, $120$ fps) suffer from noticeable spatiotemporal compression artifacts. Videos with camera motion tend to show worse quality at lower frame rates (\eg, $30$ fps), as shown in Fig.~\ref{fig:dmos_bitrate1}, where $60$ and $120$ fps converge around $12$ Mbps, with $120$ fps providing the best quality at higher bitrates (\eg, 15 Mbps).

For videos without camera motion but significant temporal variations, the results in Fig.~\ref{fig:dmos_bitrate3} suggest that $30$ fps performs better. This finding may seem counterintuitive, as one might expect higher frame rates to always enhance perceptual quality in high-motion content, given a sufficient bitrate. However, this expectation may hold only when the bitrate is above $15$ Mbps. When bandwidth is limited, increasing frame rate fails to maintain adequate spatial quality, and the resulting spatiotemporal artifacts may be amplified.

When there is minimal temporal variation, the differences between frame rates become less pronounced, as shown in Fig.~\ref{fig:dmos_bitrate2}. $30$ and $60$ fps achieve similar perceived quality at around $9$ Mbps. At higher bitrates (\eg, $15$ Mbps), however, both $60$ and $120$ fps yield similarly better quality, suggesting that even for content with minimal temporal complexity, increasing frame rate does not offer substantial perceptual benefits until a sufficient bitrate is available to provide high-quality motion representations.

\subsection{Statistical Significant Tests}
\noindent\textbf{Frame-Rate Effect.}
To assess whether frame rate affects perceived video quality, we grouped the DMOSs according to three different frame rates, and conducted a one-way analysis of variance (ANOVA). With a significance threshold set at $p=0.05$, the ANOVA results reveal a significant effect of frame rate on perceived quality ($p=4.47\times10^{-4}$).

\noindent\textbf{Joint Effect.}
Two key interactions were explored in our analysis: (1) the interaction between bitrate and frame rate and (2) the interaction between source content and frame rate. For the first interaction, we examined the effect of bitrate and frame rate in tandem, focusing on the bitrate levels at which perceptible trade-offs occur. A two-way ANOVA reveal a significant interaction ($p=4.15\times10^{-5}$), confirming that bitrate and frame rate jointly influence perceived quality, particularly at certain bitrate thresholds.
For the second interaction, we conducted a similar two-way ANOVA with source content and frame rate as independent variables and DMOS as the dependent variable. The results show a significant interaction between source content and frame rate ($p=4.79\times10^{-3}$), suggesting that the perceptual impact of frame rate varies depending on source content.

\begin{table}[t]
\captionsetup{font=small}
\small
\caption{Performance comparison of VQA models on HFR-LS.}
\label{tab:obj_models}
\centering
\begin{threeparttable} 
\begin{tabular}{l|l|cc}
\toprule[1pt]
    \multicolumn{2}{c|}{Model} & PLCC $\uparrow$ & SRCC $\uparrow$ \\
    \hline
    \multirow{5}*{FR} & PSNR & $0.112$ & $0.096$   \\
    & SSIM~\cite{wang2004ssim} & $0.307$ & $0.296$   \\
    & LPIPS~\cite{zhang2018lpips}  & $0.219$ & $0.263$  \\
    & DISTS~\cite{ding2022image} & $0.463$ & $0.356$  \\
    & VMAF~\cite{li2016VMAF}  & $0.195$ & $0.126$    \\
    \hline
    \multirow{6}*{NR} &  NIQE~\cite{mittal2013niqe}  & $0.303$  & $0.216$ \\ 
    & VSFA~\cite{li2019vsfa} & $0.508$  & $0.256$  \\
    & Li22~\cite{li2022blind} & $0.359$  & $0.275$  \\
    & DOVER~\cite{wu2023dover} & $0.582$ & $0.336$ \\
    & ModularVQA~\cite{wen2024moular} & $0.487$  & $0.380$  \\
    & MinimalisticVQA~\cite{sun2024minimalistic} & $0.596$ & $0.407$ \\
\bottomrule[1pt]
\end{tabular}
\end{threeparttable}
\end{table}

\section{VQA Model Evaluation}
To further assess the HFR-LS dataset's utility in advancing research on bitrate-constrained live streaming, we systematically evaluated several objective VQA models. The evaluation focused on two key performance metrics: PLCC, which assesses prediction linearity, and Spearman's rank order correlation coefficient (SRCC), which evaluates prediction monotonicity. For PLCC calculation, we first linearized model predictions using a four-parameter logistic function, as recommended by~\cite{video2000final}. We considered a range of full-reference VQA (FR-VQA) models, including PSNR, SSIM~\cite{wang2004ssim}, LPIPS~\cite{zhang2018lpips}, DISTS~\cite{ding2022image}, and VMAF~\cite{li2016VMAF}. To accommodate different frame rates in the dataset, we performed temporal upsampling via frame duplication to align with the reference frame rate.
Additionally, we incorporated no-reference VQA (NR-VQA) models, including NIQE~\cite{mittal2013niqe} and five learning-based ones~\cite{li2019vsfa,li2022blind,wu2023dover,wen2024moular,sun2024minimalistic}.

The results are presented in Table~\ref{tab:obj_models}, from which it is evident that FR-VQA models generally exhibit poor correlation with subjective quality scores. This underperformance is likely due to the difficulty these models face when dealing with videos at varying frame rates. MinimalisticVQA~\cite{sun2024minimalistic}, equipped with a strong temporal quality analyzer, outperforms the other models, achieving the highest (though still modest) correlation with subjective ratings.

\section{Conclusion}
We have investigated the perceptual trade-offs between compression strength and frame rate in bitrate-constrained live streaming. By creating the HFR-LS dataset, we provided valuable insights into how varying compression strengths and frame rates influence perceived video quality. Our findings reveal significant interactions between bitrate, frame rate, and source content. Importantly, increasing frame rate may amplify spatiotemporal artifacts, particularly when the bitrate is insufficient to maintain spatial quality. This highlights the need for a careful balance between compression strength and frame rate in bitrate-constrained environments.

Our results also underscore the importance of optimizing encoding parameters to achieve high-quality streaming experiences and suggest that future VQA models should incorporate a more nuanced understanding of these trade-offs. In particular, models that can account for the dynamic interactions between bitrate, frame rate, and source content will be essential for improving live video streaming systems.

\small
\section{Acknowledgments}
This work was supported by the Hong Kong ITC Innovation and Technology Fund (9440379 and 9440390).

\small
\bibliographystyle{IEEEbib}
\bibliography{ref}

\end{document}